\documentclass[aip,apl,noshowkeys,amsmath,amssymb,amsfonts,superscriptaddress,reprint]{revtex4-1}
\usepackage{graphicx}
\usepackage{bm}
\begin{document}
\title{Fast and rewritable colloidal assembly
via field synchronized particle swapping}
\author{Pietro Tierno}
\email{ptierno@ub.edu}
\affiliation{Estructura i Constituents de la Mat\`{e}ria, Universitat de Barcelona, Spain}
\affiliation{Institut de Nanoci\`{e}ncia i Nanotecnologia IN$^2$UB, Universitat de Barcelona}
\author{Tom H. Johansen}
\affiliation{Department of Physics, University of Oslo,P. O. Box 1048, Blindern, Norway}
\affiliation{Institute for Superconducting and Electronic Materials, University of Wollongong, Wollongong, NSW 2522, Australia}
\author{Thomas M. Fischer}
\affiliation{Institut f\"ur Experimentalphysik V, Universit\"at Bayreuth, 95440 Bayreuth, Germany}
\date{\today}
\begin{abstract}
We report a technique to realize reconfigurable colloidal
crystals by using the controlled motion of particle defects above
an externally modulated magnetic substrate. The transport particles is induced by applying a
uniform rotating magnetic field to a ferrite garnet
film characterized by a periodic lattice of magnetic bubbles. 
For filling factor larger than one colloid per bubble domain, the particle current arises from propagating defects where particles
synchronously exchange their position when passing from one occupied domain to the next. 
The amplitude of an applied alternating magnetic field
can be used 
to displace the excess particles via a swapping mechanism, or to mobilize the
entire colloidal system at a predefined speed.
\end{abstract}
\maketitle
The flow of electrical current in a conductor arises when electrons at the Fermi level are
scattered from occupied states into unoccupied ones due to the interaction
with an external electric field. Following this electronic analogy, it is a
compelling idea to find ways to separate an ensemble of identical colloidal
particles into a set of immobile
low energy particles, and 
colloids which can become mobile 
due to an external field.
When applied to
microtechnological devices like lab-on-a-chip,
this concept has demonstrated precise single particle operations based on
the selective motion of colloidal inclusions.~\cite{Terray2002,Sawetzki2008,Kavcic2009}\\
Paramagnetic microspheres, which can be remotely controlled via non invasive
magnetic fields,
are currently employed in biotechnological applications.~\cite{Hafeli2009,Miller2001,Graham2004}
The surfaces of these particles can be chemically functionalized
allowing to bind selectively to defined targets.~\cite{Jeong2007}
For an ensemble of monodisperse particles, the formation of a threshold energy 
where only a fraction of particles will move in
response to an applied magnetic field is
difficult to achieve.
In most cases, the field-induced interactions between the particles 
favor collective motion rather than selected displacements. 
Magnetic patterned substrates have shown considerable potential 
to overcome the above limitation. These patterns can generate strong
localized field gradients, allowing controlled particle
trapping and transport along predefined magnetic tracks.
Recent experiments include the use of arrays of permalloy elliptical
islands~\cite{Gunnarsson2005}, cobalt microcylinders,~\cite{Yellen2005} 
domain wall conduits,~\cite{Donolato2010}
magnetic wires,~\cite{Vieira2009} exchange bias layer 
systems,~\cite{Ehresmann2011} and magnetic micromoulds.~\cite{Demiros2013}\\
An alternative method consists in using epitaxially grown ferrite garnet films (FGFs),
where magnetic domains with the width of few microns, i.e. 
on the particle scale, self-assemble
into patterns of stripes or bubbles.~\cite{Seshadri1993} 
Originally developed for magnetic memory
applications~\cite{Eschenfelder1980} and magneto-optical imaging,~\cite{Johansen2004} the FGFs are
ideal to manipulate paramagnetic colloids,~\cite{Tierno2009}
or superconducting vortices.~\cite{Helseth2002}
The highly localized driving force originates
from the stray field gradient generated by the Bloch walls (BWs)
in the FGF. When properly synthesized, the displacement of the BWs
caused by an external field is smooth and reversible,
with absence of wall pinning,
thus creating a precise and controllable
driving force for the particle motion.
Here, we use a FGF to manipulate and transport an ensemble of paramagnetic
particles or a fraction of it, demonstrating a technique to
dynamically organize a colloidal system
into trapped immobile particles
and particles which become mobile above a threshold field.\\
A bismuth-substituted ferrite garnet of composition
Y$_{2.5}$Bi$_{0.5}$Fe$_{5-q}$Ga$_q$O$_{12}$
($q = 0.5-1$) 
was prepared by liquid phase epitaxial growth on a (111)-oriented
gadolinium gallium garnet (GGG) substrate.
Oxide powders of the constituent elements,
as well as PbO and B$_2$O$_3$, were initially melted at $1050 \, ^{\circ}C$ in a
platinum crucible while the GGG wafer was located horizontally just 
above the melt surface. After lowering the temperature to $700 \, ^{\circ}C$,
growth of the FGF was started by letting the substrate touch the melt. 
Keeping it there for $8$ minutes produced a FGF of $5$ micron thickness,
more details can be found in a previous work.~\cite{Helseth2001}
At ambient temperature the FGF has a spontaneous magnetization
perpendicular to the plane of the film. To minimize the magnetic energy,
the FGF breaks up into domains characterized by a labyrinth stripe pattern, 
easily observed by polarized light microscopy due to the large 
Faraday effect in this material.
High frequency magnetic fields were used to transform this pattern
into a regular triangular lattice of "magnetic bubbles". 
These are cylindrical ferromagnetic domains magnetized oppositely
to the remaining continuous area of the FGF, and separated by BWs.
The magnetic bubbles have in zero field a diameter of $6.4 \, \mu m$ and
a lattice constant of $l = 8.6 \, \mu m$, Fig.1(a).\\
To avoid particle adhesion to the film, 
the FGF was coated with a $1 \, \mu m$ thick polymer layer composed of a
positive photoresist (AZ-1512, Microchem).~\cite{Tierno22012}
Paramagnetic microspheres with $2.8 \mu m$ diameter (Dynabeads M-270)
were diluted in highly deionized water (milli-Q, Millipore), and deposited above the FGF.
After $\sim 5$ min of sedimentation, the particles became two-dimensionally confined
above the FGF due to
balance between repulsive electrostatic
interaction with the polymer layer and magnetic attraction.\\
The magnetic field applied in the $(x;z)$ plane was generated by using two custom-made
\begin{figure}[!t]
\includegraphics[width=\columnwidth]{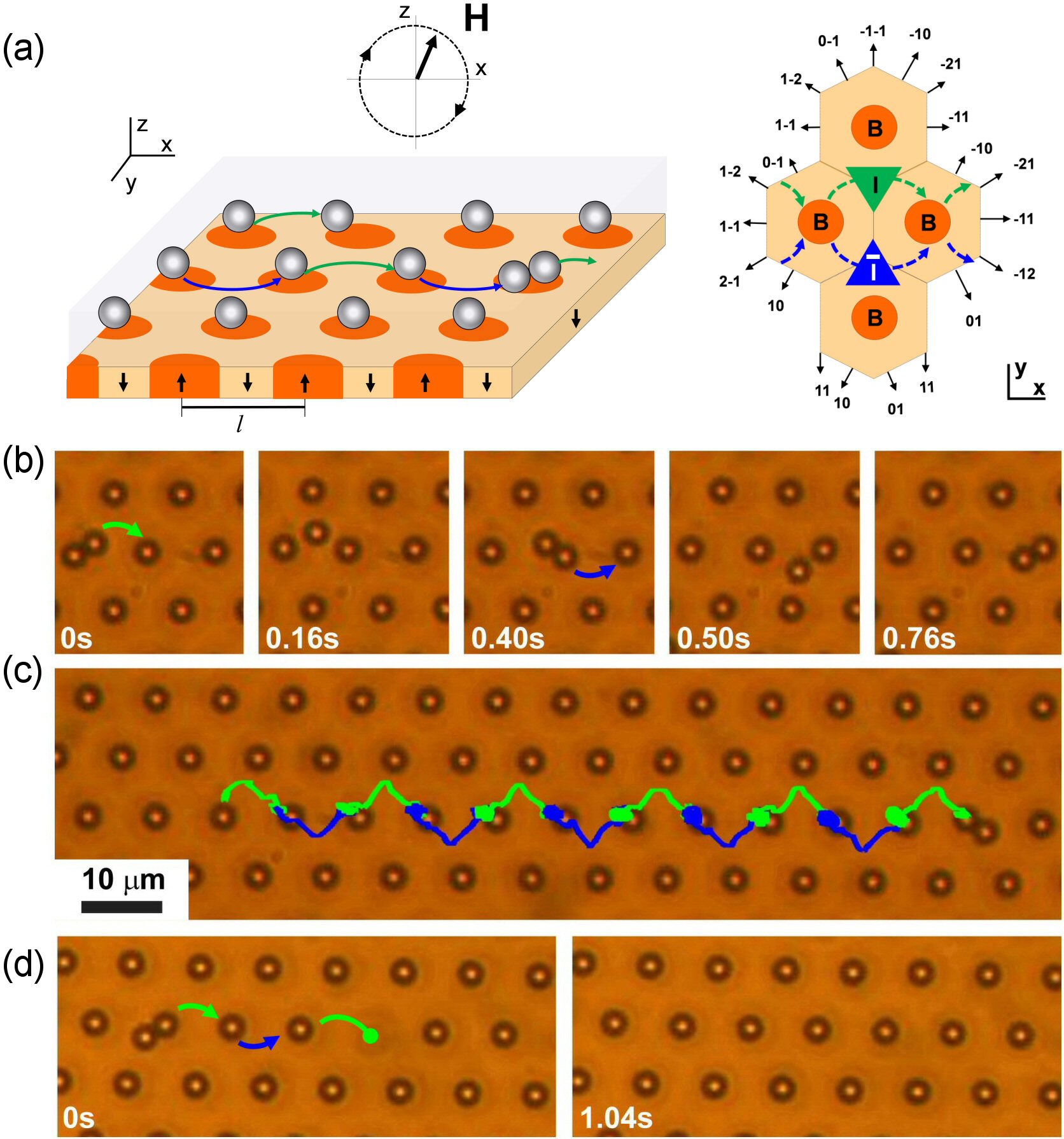}
\caption{\label{Fig_1}(a) Sketch of the magnetic bubble lattice
(lattice constant $l = 8.6 \, \mu m$) with paramagnetic colloids
and subjected to a magnetic field rotating in the $(x;z)$ plane.
Schematic on the right shows four magnetic bubbles (B) with two interstitials regions
($I,\bar{I}$) illustrating the two possible particle pathways
(dashed lines) and corresponding crystal directions.
(b) Sequence of images showing the transport via particle swapping.
(c) A particle defect transported via particle swapping, VideoS1.
Superimposed are the trajectories of the particles,
in green (blue) are trajectories along the $BIB$ ($B\bar{I}B$) pathway, VideoS1 (Multimedia View).
(d) Filling of a lattice vacancy in the colloidal crystal, VideoS2 (Multimedia View).
In these experiments, the fields have amplitudes $H_x = 0.7 \, kA/m$,
$H_z = 1.0 \, kA/m$ and frequency $\omega  = 18.8 s^{-1}$.}
\end{figure}
Helmholtz coils assembled above the stage of an upright microscope (Eclipse Ni, Nikon)
equipped with a $100 \times$ $1.3$ NA objective and a $0.45 \times$ TV lens.
The coils were arranged perpendicular to each other and connected to two independent power
amplifiers (KEPCO BOP) driven by an arbitrary waveform generator (TTi-TGA1244).\\
In the absence of an external field, the BWs of the magnetic bubbles attract the paramagnetic colloids
and without the polymer coating, the particles sediment above the BWs.
\begin{figure}[!t]
\includegraphics[width=\columnwidth]{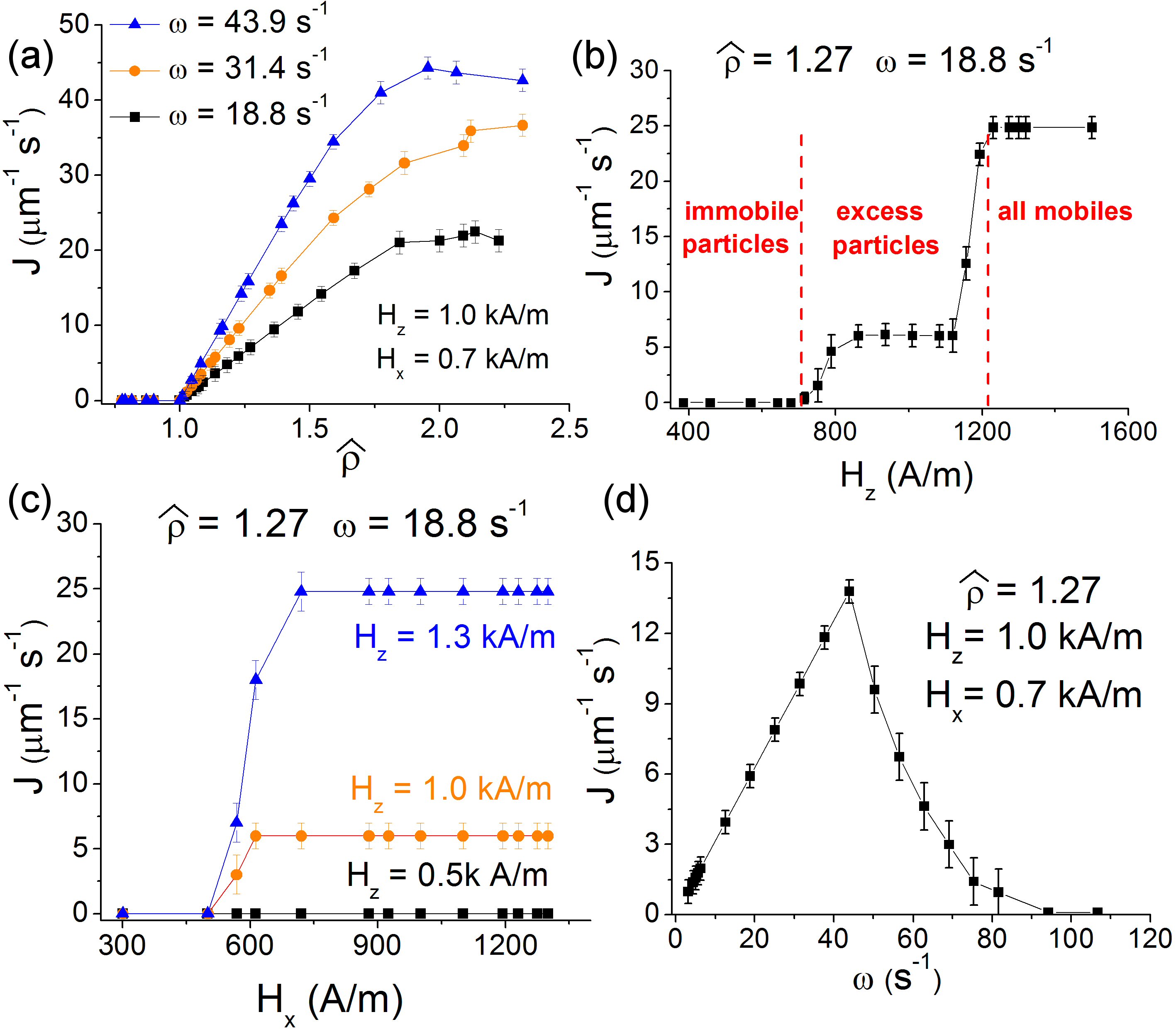}
\caption{\label{Fig_2}(a) Particle flux $J$ versus normalized density $\hat{\rho}$ for
different frequencies.
(b) $J$ versus amplitude of the perpendicular field $H_z$ ($H_x = 0.7 \, kA/m$).
Dashed red lines separate the regimes where
all particles are immobile ($H_z < 0.7 \, kA/m$), only propagation of excess particles occurs
($0.7 \, kA/m < H_z < 1.2 \, kA/m$), and all particles are mobilized ($H_z > 1.2 \, kA/m$). (c)
$J$ versus amplitude of the in-plane field $H_x$ for three different
values of $H_z$ (d) Dependence of
$J$ on the driving frequency $\omega$ for $\hat{\rho} = 1.27$.}
\end{figure}
However, due to the polymer film, the particles
have an higher elevation from the surface of the FGF
and the magnetic potential become smoother, featuring energy minima at the centers of the magnetic bubbles.~\cite{EPAPS}
Thus, once deposited above the FGF, the particles form a perfect triangular lattice for a normalized areal
density $\hat{\rho} \equiv \rho a = 1$, Fig.1. Here
$\rho = N/A$ is the particle number density, $N$ is the number of particles located within
the observation area $A=140 \times 105  \, \mu m^2$ and
$a = 64 \, \mu m^2$ is the area of the Wigner Seitz (WS) unit cell around one bubble.
For $\hat{\rho} >1$, the excess particles redistributed within
the magnetic domains, and each bubble became populated by colloidal doublets, triplets or larger clusters.\\
We induce particle motion by applying an external magnetic field rotating
in the $(x; z)$ plane, ${\bm H} \equiv (H_x \sin{(\omega t)}, 0, H_z \cos{(\omega t)})$, with angular frequency $\omega = 2 \pi \nu$
and amplitudes $(H_x;H_z)$, Fig.1(a).  In most of the experiments, we keep fixed the amplitude of
the in-plane component $H_x = 0.7 \, kA/m$, and change the ellipticity of
the applied field by varying the amplitude of the
perpendicular component, $H_z$.\\
For amplitudes  $0.8 \, kA/m < H_z < 1.2 \, kA/m$,
and loading $\hat{\rho} >1$, the particle transport
takes place via a swapping mechanism in doubly occupied domains, where adjacent particles synchronously
exchange their positions. The bubble lattice thereby preserves the overall occupancy
of one particle per domain. Increasing the particle density, the swapping motion occurs
in the form of creation/destruction of doublets, triplets or even tetramers. However, for a wide range of
particle densities, we find that colloidal defects propagate
mainly via doublets swapping motion, and the latter is illustrated in Fig.~\ref{Fig_1}(b).\\
To explain the mechanism leading to the defect propagation, let us consider the arrangement of
four bubbles with their interstitial regions, as shown in right part of Fig.~\ref{Fig_1}(a).
Energy calculations~\cite{EPAPS} show that when the rotating field becomes anti-parallel with respect to the bubble magnetization,
it generates in the interstitial regions two energy wells with triangular shape and opposite orientations,
$I$ and $\bar{I}$, with corners pointing towards the $11$ and $-1-1$ directions, respectively.
In this situation, there are two equivalent pathways along which an excess particle
can propagate towards the $-11$ direction, either along the $B\bar{I}B$ pathway (dashed blue line),
or along the $BIB$ pathway (dashed green line). Both pathways are energetically equivalent,
and the particle's choice is dictated by the initial orientation of the doublet
in the bubble domain. A doublet initially oriented along the $-21$ direction will send a
particle along the $BIB$ pathway (Fig.1(b), first three images). Afterwards, this
particle will form another doublet oriented along the $-12$ direction, which will send a
particle along the complementary $B\bar{I}B$ pathway (Fig.1(b), last three images).
When the moving particle encounters a vacancy in the colloidal lattice, it will fill it and the
defect propagation will end there, as shown in Fig.1(c).\\
We characterize the system conduction along the driving direction
by measuring the particle flux as
$J = \rho \langle v \rangle$, where $\langle v \rangle$ is the average speed as determined from particle tracking.
Fig.2(a) shows $J$ versus the dimensionless density $\hat{\rho}$, for three different
frequencies. Below the loading $\hat{\rho}=1$, i.e., having less than one particle per unit cell, there are 
no excess particles, and thus $J = 0$. For $\hat{\rho}>1$, we measure a net
colloidal current which grows linearly with the loading up to $\hat{\rho}\sim 1.6$.
In this regime, only the excess particles contribute to the current, while one particle per
magnetic bubble does not reach the mobility threshold. The speed of the particles inside the WS unit cell is
phase-locked with the driving field, and given by $v = l \omega /2 \pi $ where $l$ is the lattice constant.
Thus, increasing the driving frequency
increases the average speed and, in turn, the slope of the curve in Fig.2(a).
The direction of motion of the excess particles is 
dictated by the chirality of the rotating field,  
thus changing the polarity of one of the fields
($H_x$ or $H_z$) allows to invert the entire flow of particles across the film.\\
Increasing
the loading further ($\hat{\rho} >2$), the current reaches a maximum, and jamming between closely moving
colloids forbids further particle transport.
\begin{figure}[!t]
\includegraphics[width=\columnwidth]{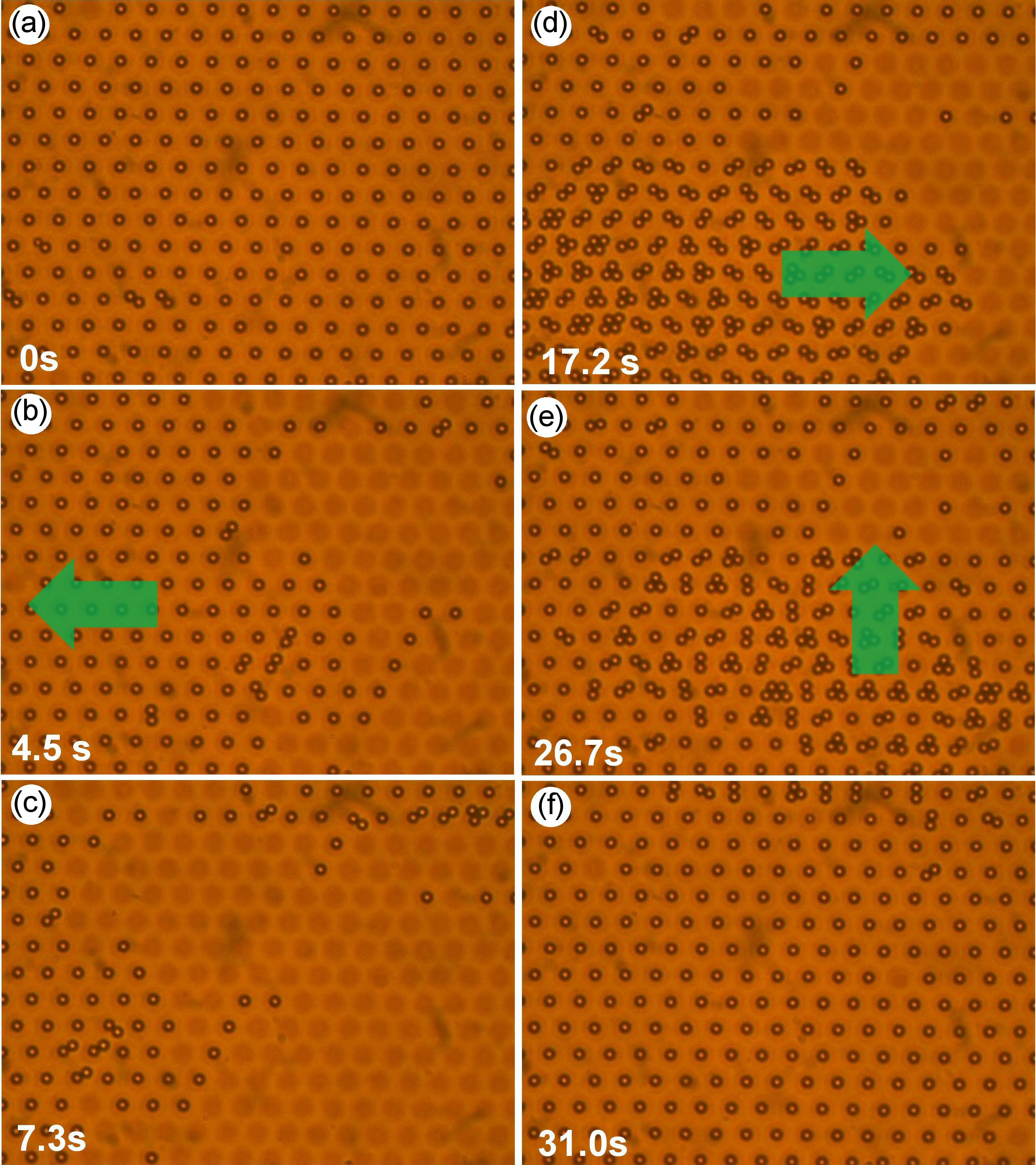}
\caption{\label{Fig_3}(a)-(f) Sequence of images
showing the reversible assembly of a colloidal lattice by switching
the applied field between the two threshold fields, $H^c_2$ and $H^c_2$. First, the
magnetic particles were driven toward the $1-1$ direction (green arrow in (b)
by an applied field rotating in the $(x;z)$ plane with components $H_x = 0.7 \, A/m$ and $H_z = 1.3 \, A/m > H^c_2$,
and frequency $\omega = 18.8 s^{-1}$.
In (d) the direction of motion is inverted and the perpendicular component of the field decreased to 
$H_z = 0.9 \, kA/m < H^c_2$.
In (e) the field rotates in the $(y; z)$ plane, and the excess particles
is transported towards the $-1-1$ direction, filling the whole
lattice of bubbles, VideoS3 (Multimedia View).}
\end{figure}
Shown in Fig.2(b) is the effect that the perpendicular component of the applied field $H_z$ has on the
particle current. This component acts directly on the size of the magnetic domains,
since the diameter of the magnetic bubbles increases (decreases) when $H_z$
is parallel (antiparallel) to the bubble magnetization. The graph, obtained for a fixed density of  $\hat{\rho} = 1.27$,
shows that the particle flux grows in discrete steps as the amplitude of $H_z$ increases.
Below $H_z = H^c_1 = 0.7 \, kA/m$, no current is observed.
Increasing $H_z$ the excess particles start to be mobilized, and the flux raises
till reaching the constant value $J = 6.1 \, \mu m^{-1} s^{-1}$ for $H_z >0.7 kA/m$.
Increasing $H_z$ further,
reveals a second field value, $H^c_2 = 1.2 \, kA/m$, where the periodic
displacements of the BWs are able to drive all particles synchronously across the film. 
The
field values $H^c_1$ and $H^c_2$  are therefore the mobility edges where different sub-ensembles of
particles can be set into motion.\\
The effect of the in-plane component of the applied field $H_x$ on the flux $J$ is shown in Fig. 2(c).
While $H_z$ controls the size of the magnetic domains, the effect of  $H_x$ is to break the spatial symmetry of the potential, 
inducing a net particle current towards a defined direction. When $H_z<H^c_1$, no current is observed for any value of
$H_x$. For $H_z = 1.0 \, kA/m$ the flux is composed of excess
particles, while for $H_z = 1.3 \, kA/m$ all particles are mobilized 
and, in both cases, when $H_x>0.7 \, kA/m$ it becomes independent on further increasing of $H_x$.\\
Fig.2(d) shows the dependence of $J$ on the driving frequency for $\hat{\rho} = 1.27$. The current displays a
full participation of the excess particles up to $\omega  \sim 44 \, s^{-1}$, where the flux reaches 
its maximum value, $J=13.8 \, \mu m^{-1} s^{-1}$ corresponding to an average defect speed of $v = 60 \, \mu m s^{-1}$.
Beyond this frequency, the participation of excess particles becomes partial, and the current decreases monotonously,
reaching zero near $\omega  = 110 \, s^{-1}$. In this second regime, the high frequency motion of
the particles is found to be intermittent, with the excess particles randomly switching between
being immobile for some period, for afterwards becoming fully mobile again,
reducing the efficiency of our magnetic device.\\
Controlling the motion of the excess particles makes it possible to easily create
or destroy a colloidal lattice by switching the applied field between the two mobility thresholds.
This concept is demonstrated in Fig.3 (Video3). Starting from a triangular
arrangement with one particle per WS cell (a) ($t = 0s$), we apply an elliptically 
polarized magnetic field with components $H_x = 0.7 \, kA/m$, $H_z = 1.3 \, kA/m > H^c_2$,
and angular frequency
$\omega  = 18.8 \, s^{-1}$, which mobilize the whole lattice at a speed of $25.8 \, \mu m s^{-1}$ towards the $1-1$ direction, 
Figs.3(b,c), leaving the bubble substrate almost unfilled. Since the moving particles 
are phase-locked with the driving field for the used frequency, 
the translating lattice is stable and preserves the initial 
triangular order during motion.   
After $t \sim 10 s$, we change the polarity of the in-plane field ($H_x \rightarrow -H_x$),
in order to invert the particle flux towards the $-11$ direction.
We also decrease the field amplitude to $H_z = 0.9 \, kA/m < H^c_2$, inducing a current composed 
only of excess particles which start filling again the bubble lattice in the bottom part of the film, 
Figs.3(d,e). 
Since, during these operations the top part of the film is left unfilled,
we change the orientation of the applied field after $t = 26.7 s$. The field now rotates 
in the $(y; z)$ plane,
and transports the excess particles toward the $-1-1$ direction, 
reforming the colloidal crystal after $31 s$.  The colloidal assembly demonstrated here can be further optimized by either
controlling the particle density, and thus the amount of excess particles propagating through the lattice,
or by increasing the driving frequency, up to a maximum speed of $v = 60 \, \mu m s^{-1}$, which will
further speed up the re-writing process.\\
In conclusion, we demonstrate a technique to remotely generate and control 
the motion of defects in two
dimensional lattices, while keeping track of the position of the individual
particles, which could be used as model system to study the dynamics of
impurities in crystalline materials. The energy scales involved in our system ($\sim 150 k_B T$, with $T\sim 293K$) are much
beyond the effect of thermal fluctuations, which could interfere with the colloidal
transport process. This makes our magnetic device fully controllable, ensuring an
extremely precise tuning of the particle speed and dynamics in real time and space.
Although our experiments focus on using FGF films as functional platform, the technique 
reported here should be applicable, within the constraint of colloidal particle size 
and lattice wavelength, to other platforms where magnetic patterns are created by 
"top-down" fabrication processes.\\
P. T. acknowledges support from the ERC starting grant "DynaMO" 
(No. 335040) and from the programs No. RYC-2011-07605, FIS2011-15948-E.
T. H. J. thanks the Research Council of Norway. T. M. F.
acknowledges support by the DFG via the center of excellence SFB 840.

\end{document}